\documentclass{elsart}
\usepackage{ifpdf}
\usepackage{graphicx,amssymb,lineno}
\ifpdf
\usepackage[%
  pdftitle={Instructions for use of the document class
    elsart},%
  pdfauthor={Simon Pepping},%
  pdfsubject={The preprint document class elsart},%
  pdfkeywords={instructions for use, elsart, document class},%
  pdfstartview=FitH,%
  bookmarks=true,%
  bookmarksopen=true,%
  breaklinks=true,%
  colorlinks=true,%
  linkcolor=blue,anchorcolor=blue,%
  citecolor=blue,filecolor=blue,%
  menucolor=blue,pagecolor=blue,%
  urlcolor=blue]{hyperref}
\else
\usepackage[%
  breaklinks=true,%
  colorlinks=true,%
  linkcolor=blue,anchorcolor=blue,%
  citecolor=blue,filecolor=blue,%
  menucolor=blue,pagecolor=blue,%
  urlcolor=blue]{hyperref}
\fi

\makeatletter
\def\elsartstyle{%
    \def\normalsize{\@setfontsize\normalsize\@xiipt{14.5}}
    \def\small{\@setfontsize\small\@xipt{13.6}}
    \let\footnotesize=\small
    \def\large{\@setfontsize\large\@xivpt{18}}
    \def\Large{\@setfontsize\Large\@xviipt{22}}
    \skip\@mpfootins = 18\p@ \@plus 2\p@
    \normalsize
}
\@ifundefined{square}{}{}
\makeatother

\graphicspath{{/home/olive/RECHERCHE/ACOUSTIQUE_URBAINE/ARTICLE/Ap_Acoustics/Figures/}}

\begin{document}

\begin{frontmatter}
\title{Effect of the open roof on low frequency acoustic propagation in street canyons}

\author{O. Richoux, C. Ayrault, A. Pelat, S. F\'elix and B. Lihoreau}
\address{LAUM, CNRS, Universit\'e du Maine, Av. O. Messiaen, 72085 Le Mans, FRANCE.}

\ead{olivier.richoux@univ-lemans.fr}
\ead[url]{}

\begin{abstract}
This paper presents an experimental, numerical and analytical study of the effect of open roof on acoustic propagation along a $3$D urban canyon. The experimental study is led by means of a street scale model. The numerical results are performed with a $2$D Finite Difference in Time Domain approach adapted to take into account the acoustic radiation losses due to the street open roof. An analytical model, based on the modal decomposition of the pressure field in a horizontal plane mixed with a $2$D image sources model to describe the attenuation along the street, is also proposed. Results are given for several frequencies in the low frequency domain ($1000-2500$ Hz). The comparison of the three approaches shows a good agreement until $f=100$ Hz at full scale, the analytical model and the $2$D numerical simulation adapted to $3$D permit to modelize the acoustic propagation along a street. For higher frequency, experimental results show that the leakeage, due to the street open roof, is not anymore uniformly distributed on all modes of the street. The notion of leaky modes must be introduced to modelize the acoustic propagation in a street canyon.
\end{abstract}

\end{frontmatter}
%%%%%%%%%%%%%%%%%%%%%%%%%%%%%%%%%%%%%%%%%%%%%%%%%%%%%%%%%%%%%%
\section{Introduction}
\label{intro}

Urban acoustic researches are divided in three thematics : the sources characterization and identification, the acoustic propagation and the noise perception in an urban context. This work enters in the second thematic : its purpose is to describe the propagation of sound emitted by known sources in a street by taking into account its physical morphology (open roof, height and width of the street).

To study acoustic propagation in an urban context, several approaches are available : energetical methods \cite{Wiener65,Oldham94,Picaut99a,Picaut99b} based on the estimation of a quadratic quantity (energy density or acoustic intensity), numerical methods \cite{Renterghm06,Albert05} to estimate acoustic pressure or velocity and modal approach \cite{Bullen76} to calculate the pressure or velocity fields.\\
Energetical methods are largely used in urban acoustics but concern a limited frequency range : the image sources method \cite{Wiener65,Oldham94,Wu95,Lu02,Kang00}, the ray tracing approach \cite{Bradley77}, the radiosity method \cite{Kang02,LeBot02,Kang05} and finally statistical methods of particle transport \cite{Picaut99a,Picaut99b,LePolles04} are generally used for middle and high frequencies. All these approaches propose to modelize the effect of the street open roof by a complete absorption of the wave.\\
The numerical methods, as Finite Element Method (FEM) or Boundary Element Method (BEM), are restricted for urban acoustics to low frequencies for $2$D case or very low frequencies for $3$D case \cite{Albert05,Renterghm06} because of the large time cost dur to the discretization.\\
The modal approach, where the geometrical characteristics of the street are explicitely taken into account in the model, is generally not used due to the complexity of the medium and to the difficulty to determine the modes of an open space like a street canyon. For example, Bullen \textit{and al} \cite{Bullen76} have studied the acoustic propagation in a guide with infinite height or more recently, the modal approach was used to calculate the $2$D field in a street section, the acoustic radiation conditions being described by an equivalent sources distribution at the interface between the street and the free space \cite{Ogren04}.

This review highlights more particularly that the open aperture of the street on the half free space, essential characteristic of the urban environment, is taken into account only in $2$D (in a section) and rarely in $3$D urban acoustic problem \cite{Heimann07}. Only the energetical approaches modelize this characteristic by a complete absorption of the wave but these methods are restricted to middle and high frequencies and this asumption of complete absorption should be justified at these frequencies and moreover at low frequencuencies.\\

The aim of this paper is to study the acoustic propagation along a $3$D street canyon and to propose models to describe the role of the street open roof in the propagation for low frequencies. This work compares experimentally, analytically and numerically the study of the open roof effect on acoustic propagation. Experimental results, obtained with a street scale model are described in section \ref{experimental}. The analytical study, based on the modal decomposition of the field in the horizontal plane mixed with an image sources model to determine the attenuation due to the street open roof is presented in section \ref{model_mode}. The numerical study, made with a $2$D-FDTD computation adapted to take into account the acoustic radiation losses by the street open roof is proposed in section \ref{numerical}. Then, in section \ref{results}, experimental, analytical and numerical pressure field maps are compared for several frequencies providing validity and limits of the models.

%%%%%%%%%%%%%%%%%%%%%%%%%%%%%%%%%%%%%%%%%%%%%%%%%%%%%%%%%%%%%%
\section{Experimental study of the open roof effect on the acoustic propagation in a street canyon}
\label{experimental}

\subsection{Experimental set-up}
\subsubsection{Scale model of the street}

\begin{figure}[h!]
\begin{center}
\includegraphics[width=15cm]{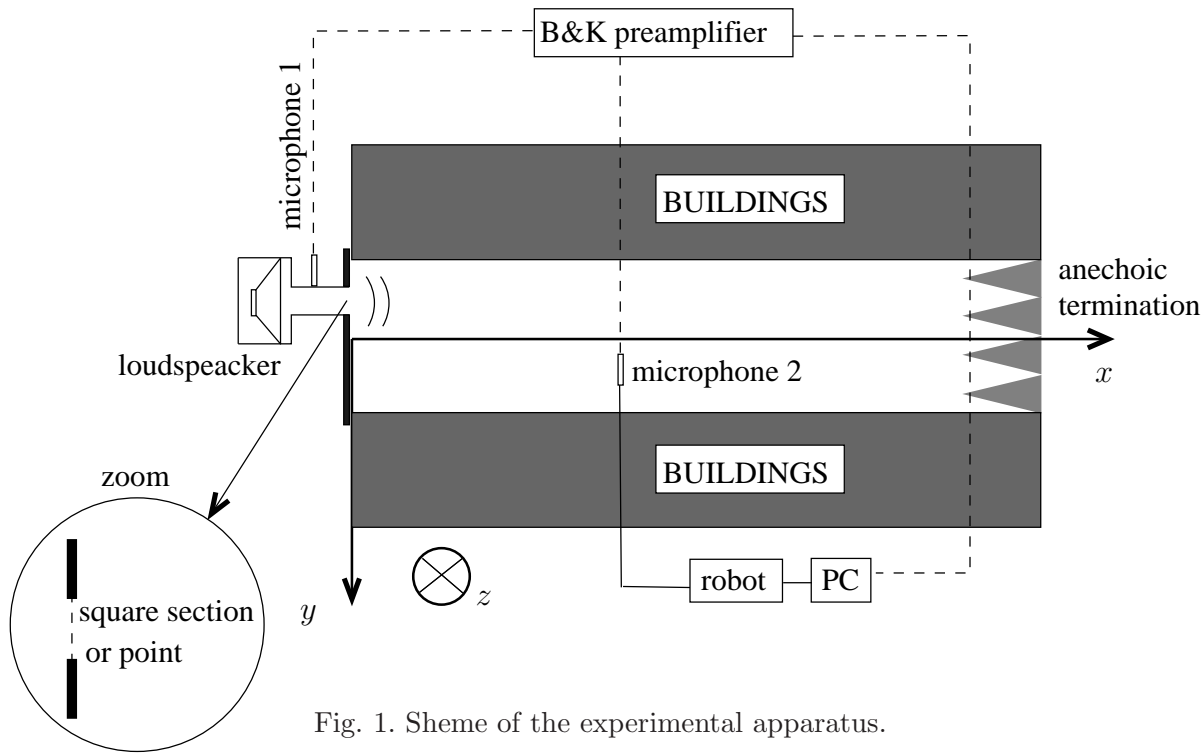}
\end{center}
\caption{Sheme of the experimental apparatus.}
\label{fig:photo_street1}
\end{figure}

A $1/25$ scale model of $0.27$ m height and $0.2$ m width with $3$ m length is carried out (see fig. \ref{fig:photo_street1}) corresponding in full scale to a $6.75$ m heigth, $5$ m width and $75$ m length street. This scale model is fully described in \cite{Picaut01} where the building facades are simulated by wood cubes with a plane distribution. The scale model is put in a semi-anechoic room.

\begin{figure}[h!]
\begin{center}
\includegraphics[width=5cm]{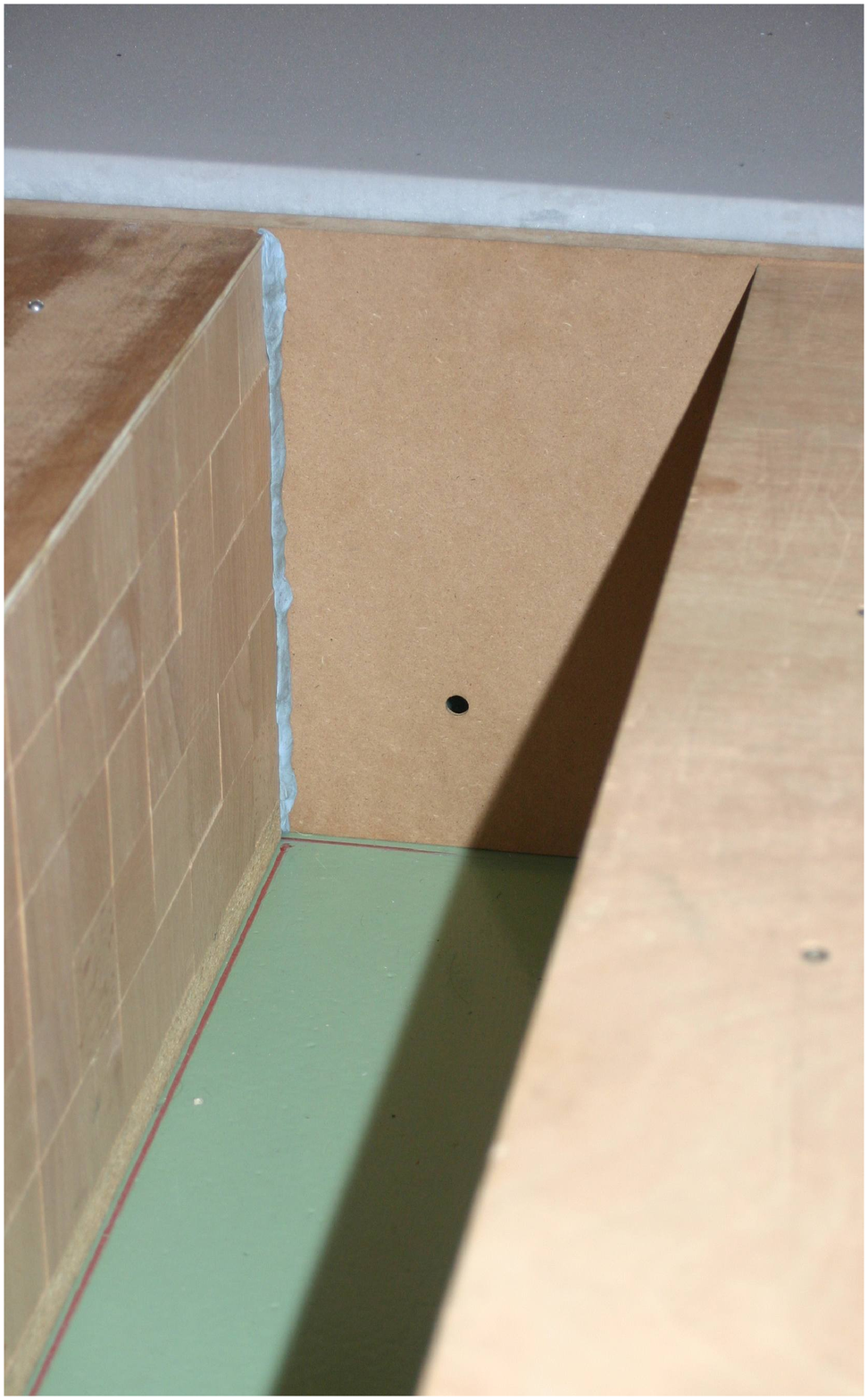}
\includegraphics[width=5cm]{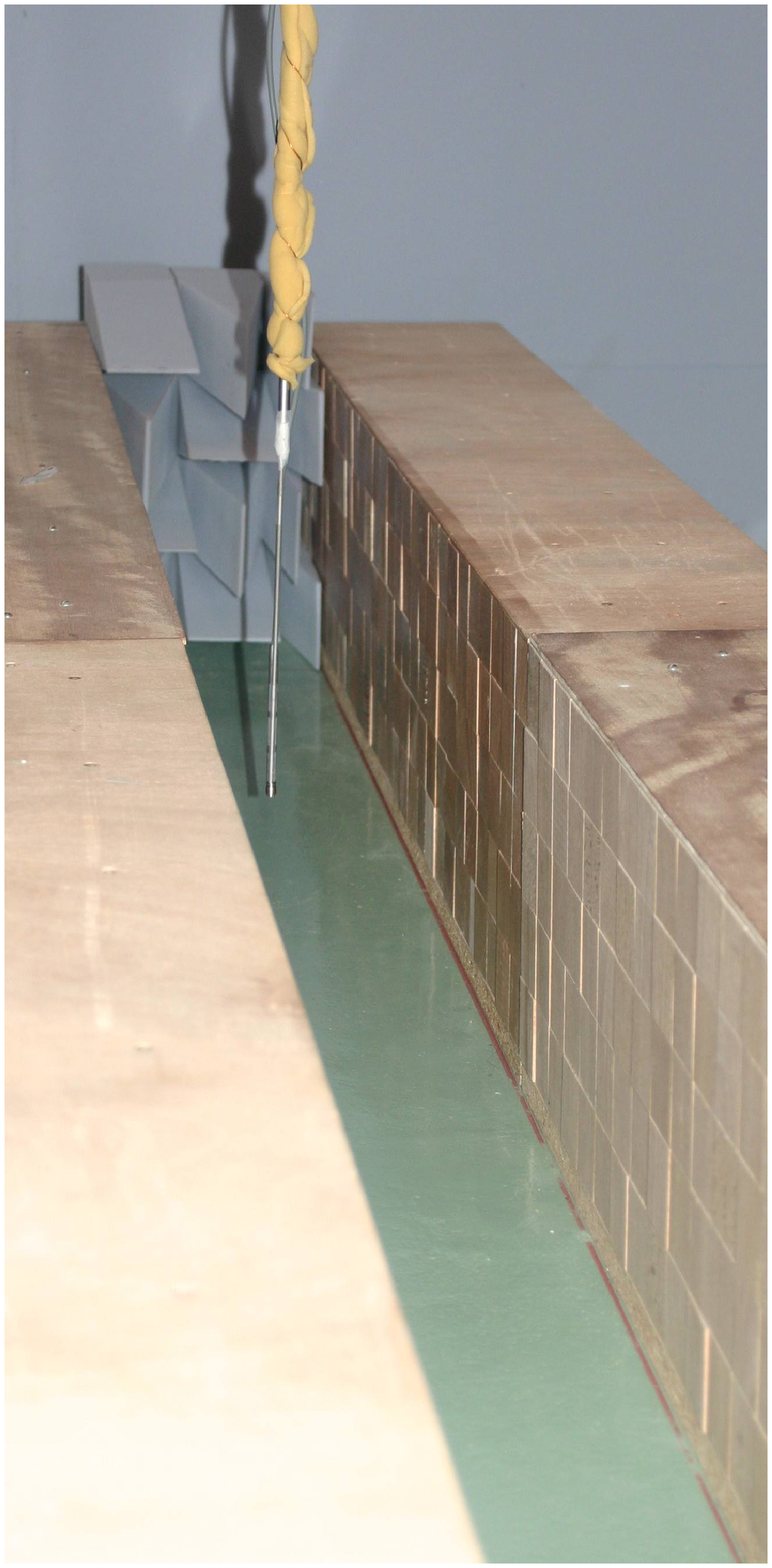}
\end{center}
\caption{Views of the street scale model. (a) Rigid wall with a circular aperture at the entry of the street scale model. (b) Street scale model with the anechoic end.}
\label{fig:photo_street}
\end{figure}

On one side, the source is enclosed (see figure \ref{fig:photo_street1}) in a flat rigid wall while an anechoic termination is carried out on the other side (fig. \ref{fig:photo_street}b). This anechoic termination is made of melamine dihedron designed to obtain a cut-off frequency around $750$ Hz. This allows to consider the scale model of $3$ m length as semi infinite for acoustic frequency upper than 1kHz.

\subsubsection{Sound source}

The acoustic source in the street is a loudspeaker enclosed in a rigid box opened on a guide with a square cross-section ($0.05 \times 0.05$ m$^2$). Two sorts of aperture in the entry rigid wall are used for the experimental studies : a $0.05 \times 0.05$ m$^2$ square cross-section aperture and a circular aperture with $0.01$ m diameter (see the fig. \ref{fig:photo_street}a). The square cross-section aperture simulates a plane source until $3400$ Hz and the circular source simulates a point source.

\subsubsection{Data acquisition system and post-processing}

The acoustic pressure is measured in the speaker box and in the street by means of two $1/4$ in. pressure microphones (B\&K 4938) connected to a preamplifier (B\&K 2670) and a conditionning amplifier (B\&K Nexus 2693). The preamplifier with the microphone is put vertically in the scale model to minimize the acoustic diffraction at high frequencies and for practical use during displacement.
The excitation signal is sinusoidal with a variable frequency $f$. The acquisition of the acoustic pressure is performed using a sampling frequency $F_s$ and during a time length $T_a$. The RMS value of the acoustic pressure is determined for each point of the defined scale model mesh. By means of a Charlyrobot $3$D-robot, a map of the RMS acoustic pressure is then available.

\subsection{Experimental results}

In this work, the acoustic propagation along a street canyon is led for low frequency case. The study of the pressure field in a horizontal plane along the street is made for the frequencies $f=1000$, $1500$, $2000$, and $2500$ Hz. This frequencies correspond to $f=40$, $60$, $80$, and $100$ Hz in full scale for a street of $6.75$ m height, $5$ m width and $75$ m length. The scale model of the street is $L=2.8$ m length, $d=0.2$ m width and $h=0.27$ m heigth.
The $3$D robot allows to obtain horizontal maps of the acoustic pressure RMS value along the street. The step of spatial sampling is $0.01$ m on the $x$-axis and $y$-axis. The sampling frequency $F_s$ is chosen as $F_s=20f$ and the acquisition time of the acoustic signal for each point of the map is defined by $T_a=N_s/F_s$ with $N_s$ the samples number (typically $N_s=2000$ which provides $100$ samples per acoustic period). The RMS value of the acoustic pressure is determined by means of a Matlab program using a least mean square method to determine the mean value, the amplitude and the phase of the signal. The normalized (to the entry) acoustic pressure RMS value (acoustic pressure in the following) along the street will be shown.

\begin{figure}[h!]
\begin{center}
\includegraphics[width=15cm]{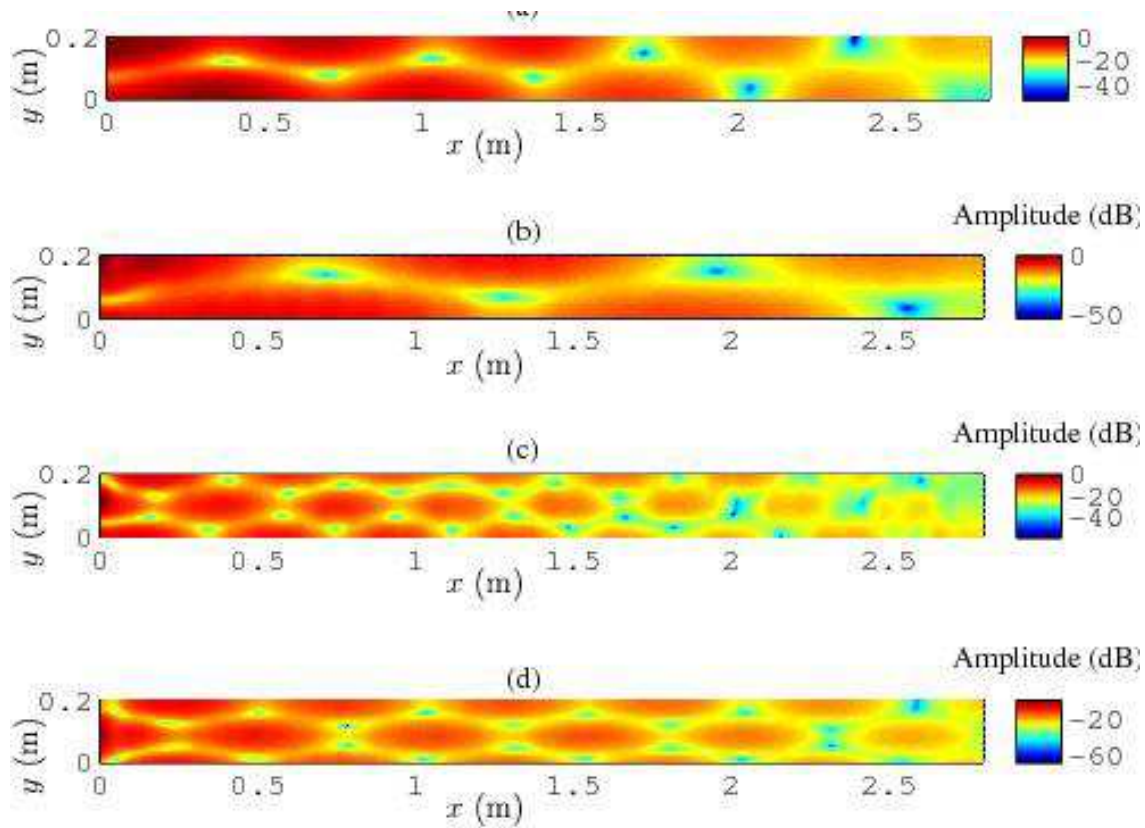}
\end{center}
\caption{Acoustic pressure in the street (scale model) for (a) $f=1000$ Hz at $z=0.07$ m and for a square source centered on $y_s=0.175$ m and $z_s=0.07$ m,  (b) $f=1500$ Hz at $z=0.07$ m and for a square source centered on $y_s=0.175$ m and $z_s=0.07$ m, (c) $f=2000$ Hz at $z=0.07$ m and for a point source centered on $y_s=0.115$ cm and $z_s=0.07$ m, (d) $f=2500$ Hz at $z=0.07$ cm and for a square source centered on $y_s=0.1$ m and $z_s=0.07$.}
\label{fig:mesure}
\end{figure}

The fig. \ref{fig:mesure}a shows the acoustic pressure map along the street for $f=1000$ Hz, with a square source centered on $y_s=0.175$ m and $z_s=0.07$ m. The map is measured at the height $z=0.07$ m. Firstly, the attenuation increases along the street. This phenomenon shows that the open roof of the street brings acoustic losses for the pressure field inside the street. This attenuation is studied more precisely in the following section. Secondly, the shape of the acoustic map shows the presence of acoustic modes in the $(x,y)$ plane of the street. For $f=1000$ Hz, this pressure map can be compared to a map of the acoustic propagation in a $2$D infinite waveguide with two propagative modes without attenuation.

The fig. \ref{fig:mesure}b shows acoustic pressure along the street at $z=0.07$ m for $f=1500$ Hz, with a square source centered on $y_s=0.175$ m and $z_s=0.07$ m. The same remarks as for $f=1000$ Hz can be made, except that the shape of the pressure map is not the same due to a different repartition of the source condition on the modes. We can note that attenuation depends on frequency showing that the acoustic losses due to the open roof depends also on frequency. The fig. \ref{fig:mesure}c shows the acoustic pressure map along the street at $z=0.07$ m for $f=2000$ Hz, with a point source centered on $y_s=0.115$ m and $z_s=0.07$ m. The modal repartition of the source condition and the number of the propagative modes differ from the two previous cases (for $f=2000$ Hz, $3$ modes are propagative). The fig. \ref{fig:mesure}d shows the acoustic pressure attenuation along the street at $z=0.07$ m for $f=2500$ Hz, with a point source centered on $y_s=0.1$ m and $z_s=0.07$. 

All these experimental maps show that the pressure on the $(x,y)$ plane can be easily modelized by a modal decomposition of the field. On the second hand, for all cases, it appears that the attenuation increases along the street and depends on frequency : the open roof of a street canyon provides acoustic losses for the acoustic fied in the street. In regards to the experiment conclusions, two modelizations of the acoustic propagation along a street are proposed in the following section : an analytical model and a numerical simulation with a description of the attenuation due to the open roof.

%%%%%%%%%%%%%%%%%%%%%%%%%%%%%%%%%%%%%%%%%%%%%%%%%%%%%%%%%%%%%%%%%%%%%%%%%%%%%%%%%%
\section{Analytical modelization of the open roof effect on acoustic propagation along a street}
\label{model_mode}

The analytical model of the acoustic propagation along a street is based on the modal decomposition of the pressure field in the horizontal plane $(x,y)$ (see fig. \ref{fig:street3}a). The attenuation of the pressure along a street, due to acoustic radiation losses through the open roof is described by means of a $2$D model propagation in free field in the vertical plane $(x,z)$ (see fig. \ref{fig:street3}b). The combination of these two approaches allows to elaborate a $3$D analytical model of the acoustic propagation along a street.

\subsection{Acoustic propagation in a $2$D waveguide}

\begin{figure}[h!]
\begin{center}
\includegraphics[width=14cm]{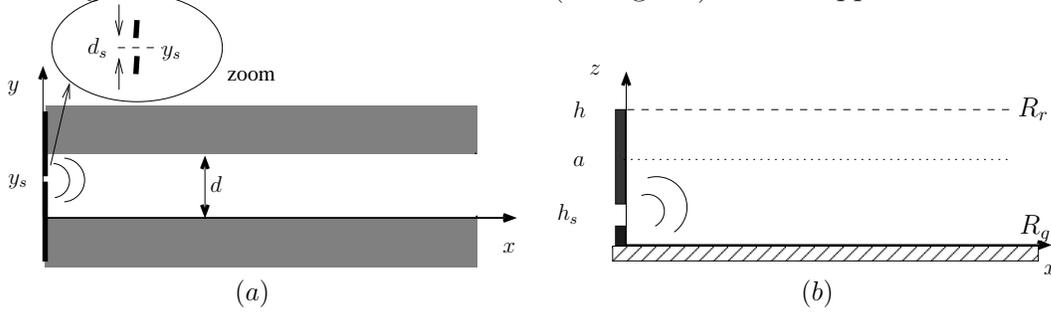}
\end{center}
\caption{View of the street in the $(x,y)$ plane (a) and  in the $(x,z)$ plane (b).}
\label{fig:street3}
\end{figure}

A semi-infinite $2$D wave guide of width $d$ closed by a rigid wall at $x=0$ containing an acoustic source is considered (see fig. \ref{fig:street3}a).
In the approximation of the linear acoustic and in the case of a perfect fluid, the acoustic pressure $p(t)$ and the acoustic velocity $\vec{v}(t)$ satisfy the equations of the mass conservation law
\begin{equation}
\frac{dp(t)}{dt} + c_0^2 \rho_0 \vec \nabla . \vec{v}(t)=0,
\label{mass_conservation1}
\end{equation}
and of the impulsion conservation (Euler equation)
\begin{equation}
\rho_0 \frac{d\vec{v}(t)}{dt} + \vec{\nabla} p(t)=\vec{0},
\label{Euler1}
\end{equation}
where $c_0$ is the sound celerity and $\rho_0$ is the air density. In the frequency domain (with a temporal convention $e^{j \omega t}$, $\omega$ being the acoustic pulsation), eqs. (\ref{mass_conservation1}) and (\ref{Euler1}) allow to write the propagation problem, outside the acoustic source and with rigid boundaries, under the following form
\begin{eqnarray}
(\Delta + k^2) p=0, \label{eq:propagation1}\\
\frac{\partial p}{ \partial n} = 0,
\label{eq:propagation2}
\end{eqnarray}
where $p$ is here complex variable, $k=\omega/c_0$ is the wave number and $\partial /\partial n$ is the normal wall derivative. The solution of the acoustic problem given by eqs. (\ref{eq:propagation1}) and (\ref{eq:propagation2}) is written as
\begin{eqnarray}
p(x,y)= \sum_{n=1}^N A_n \sqrt{2-\delta_{n0}} \cos(\frac{n \pi}{d} y) e^{-j k_{nx} x},
\label{eq:sol_pression}
\end{eqnarray}
where $k^2_{nx}=k^2 - (n \pi /\d)^2$, $\delta_{n0}$ is the Kronecker symbol ($\delta_{n0}=\delta(n) = 1$ for $n=0$ and $\delta_{n0} =0 \, \forall n \neq 0$) and $N$ is the number of modes. To determine the amplitude $A_n$ of each mode, the boundary condition given by the acoustic source is used. We suppose that the pressure field is calculated in the horizontal plane containing the source. The source is represented by a hole with a width $d_s$ in a rigid wall on the street entry and located at the point $(0,y_s)$ belonging to the $(x,y)$ plane (see fig. \ref{fig:street3}a). The source condition is described in the following form

\begin{eqnarray}
& v(y)=1, \mbox{ for } y_s-d_s/2 < y < y_s+d_s/2, \label{eq:vitesse1} \\ 
& v(y)=0, \mbox{ for } 0 < y \leq y_s-d_s/2 \mbox{ and } y_s+d_s/2 \leq y < d.
\label{eq:vitesse2}
\end{eqnarray}
The projection of the source conditions given by eqs. (\ref{eq:vitesse1}) and (\ref{eq:vitesse2}) on the modal basis allows to determine the modal amplitude of each mode $A_n$.

\subsection{Determination of the attenuation due to the open roof}

To modelize the attenuation of the pressure field due to acoustic radiation losses through the street open roof, a $2$D propagation model is used in the vertical plane $(x,z)$ (see fig. \ref{fig:street3}b). In this plane, the ground of the street is considered as perfectly rigid with a reflexion coefficient $R_g=1$ and the source is described by a point source located at the height $h_s$ embedded in a rigid wall with height $h$. The acoustic radiation condition describing the open roof of the street is modelized by means of a reflexion coefficient $R_r$ at the height $h$ on the $z$-axis.

The pressure field in the street can be decomposed as an infinite sum describing the multiple reflexions on the ground and on the street roof. In a $2$D domain, the Green function $G(\vec{r},\vec{r}_0)$ is written as 
\begin{equation}
G(\vec{r},\vec{r}_0)=-\frac{j}{4}(H_0^1(k|\vec{r}-\vec{r}_0|)),
\label{eq:Green_2D}
\end{equation}
where $\vec{r}_0$ defines the position of the source and $H_0^1$ is the Hankel function of first order. According to eq. (\ref{eq:Green_2D}), the pressure field, at the altitude $z$ in the street, takes the following form

\begin{eqnarray}
p(x,z)=-\frac{j}{4}\sum_{m=-\infty}^{+\infty}(R_r)^{|m|} & \left[ H_0^1 \left( k\sqrt{x^2+(z-2mh-h_s)^2} \right) \right. \nonumber\\
& +H_0^1 \left. \left( k\sqrt{x^2+(z+2mh+h_s)^2} \right) \right],
\label{eq:source_image}
\end{eqnarray}
where $m$ informs on the number of image sources used in the model.

\subsection{Analytical model of the acoustic propagation in a street}

The complete modelization of the pressure field along the street is established into two steps :
\begin{enumerate}
\item the first step consists to calculate the pressure field along the street in the horizontal plane $(x,y)$ without any attenuation by means of the eq. (\ref{eq:sol_pression}) and after having determined the modal amplitude of each mode by using the source conditions described by the eqs. (\ref{eq:vitesse1}) and (\ref{eq:vitesse2});
\item in the second step, the attenuation of the pressure field for an altitude $z$, described by the eq. (\ref{eq:source_image}) is applied to each point of the pressure field in the horizontal plane $(x,y)$. For this calculus, the height of the source $h_s$ and the altitude $z$ of the horizontal plane in which the pressure field is determined are necessary. 
\end{enumerate}
Finally, the pressure field along the street, at the altitude $z$, taking into account the acoustic radiation losses due to the open roof is given by the following equation :
\begin{eqnarray}
& p(x,y,z)= -\frac{j}{4} \sum_n \sum_m (R_r)^{|m|} \left\lbrace A_n \sqrt{2-\delta_{n0}} \cos(\frac{n \pi}{d} y) e^{-j k_{nx} x} \right. \nonumber \\
& \left. \left[ H_0^1 \left( k\sqrt{x^2+(z-2mh-h_s)^2} \right) +H_0^1 \left( k\sqrt{x^2+(z+2mh+h_s)^2} \right) \right] \right\rbrace .
\label{eq:pressure_total}
\end{eqnarray}

In this work, the reflexion coefficient $R_r$ describing acoustic radiation leakage due to the street open roof is determined by fitting the modelization with the experimental results obtained with the street scale model. 

%%%%%%%%%%%%%%%%%%%%%%%%%%%%%%%%%%%%%%%%%%%%%%%%%%%%%%%%%%%%%%%%%%%%%%%%%%%%%
\section{Numerical simulation of the open roof effect on acoustic propagation along a street}
\label{numerical}

The acoustic pressure field along the street with a height $h$ is simulated using a two-dimensional finite difference time domain (FDTD) computation modified to take into account the acoustic radiation losses through the street open roof. 

To obtain this adaptated FDTD model, a $3$D-description of the acoustic propagation is reduced to an equivalent $2$D description. In this $2$D model, the attenuation (called also leakage) due to the third dimension (here the $z$-axis) is introduced by means of negative source term $q(t)$  in the mass conservation law (eq. (\ref{mass_conservation1})) leading to the following form 
\begin{equation}
\frac{dp(t)}{dt} + c_0^2 \rho_0 \vec\nabla . \vec{v}(t)= \rho_0 c_0^2 q(t).
\label{eq:mass_conservation3}
\end{equation}
This negative source, uniformly distributed in the $2$D plan, depends on the pressure $p(t)$ and can be written as
\begin{equation}
q(t) = - \alpha p(t),
\label{eq:leak}
\end{equation}
where $\alpha > 0$ is a coefficient to be determined. This source term can be considered as the effect of the third dimension (attenuation or leakage) on the propagation of the wave in the $2$D space $(x,y)$.

The discretized forms of the eqs. (\ref{eq:mass_conservation3}), (\ref{eq:leak}) and the Euler equation given by the eq. (\ref{Euler1}) can be achieved by a two dimensional finite difference time domain (FDTD) computation. After the integration of the mass conservation along a surface element of dimension $dx$ along the $x$-axis and $dy$ along the $y$-axis,  we obtain  
\begin{equation}
\frac{dp}{dt}(x,y)+ c_0^2 \rho_0 \left( \frac{dv_x}{dx}(x,y) + \frac{dv_y}{dy}(x,y) \right) = -\rho_0 c_0^2 \alpha p(x,y),
\label{mass_conservation_4}
\end{equation}
where $v_x$ and $v_y$ are respectively the projectionss of the acoustic velocity along the $x$ and $y$ axis. The same approach is used with the Euler equation (eq. (\ref{Euler1})) and a temporal discretization of these two equations allows to obtain a $2$D-FDTD computation \cite{Redondo07} of the propagation in a $3$D space.

To apply the adaptated $2$D-FDTD simulation to the propagation along a street with an open roof, the discretized equations are computed by means of a Matlab program where the boundary conditions are introduced on the acoustic velocity (the facades of the street are considered as perfectly rigid) and the source condition is introduced on the pressure. The coefficient $\alpha$ in the equation (\ref{mass_conservation_4}) can be considered as an attenuation for the $2$D propagation in the $(x,y)$ plane and describes the acoustic leakage due to the open roof of the street. This coefficient depends on the altitude $z$ of the $2$D plan $(x,y)$ in comparison with the source height $h_s$ in the street, on the aspect ratio of the street $d/h$ and on the acoustic radiation condition on the street roof.

%%%%%%%%%%%%%%%%%%%%%%%%%%%%%%%%%%%%%%%%%%%%%%%%%%%%%%%%%%%%%%%%%%%%%%%%%%%
\section{Results and discussion}
\label{results}

%-----------------------------------------------------------------------
\subsection{General observations}

In this section, we propose a qualitative comparison between the experimental, analytical and numerical results for $f=1000$ Hz, $f=1500$ Hz and $f=2500$ Hz. 

Two simulated results of the acoustic pressure field in a street with the adapted-$2$D FDTD computation are proposed. The simulation modelizes the experimental source condition and a Perfect Matching Layers (PML) is used to describe the anechoic terminaison of the street. The comparison with the experimental results is led here for $f=1000$ Hz and $f=2500$ Hz. The coefficient $\alpha$ modelizing the acoustic losses due to the street open roof is adapted by means of a minimization algorithm with the experimental data.

The results of the analytical model are proposed for $f=1000$ Hz and $f=1500$ Hz. The acoustic radiation conditions of the open roof are described by a reflexion coefficient in the model. As planned, the reflexion coefficient is negative and its value is adapted by fitting the analytical results with the experimental ones. The analytical field is calculated by means of a modal decomposition with $N$ modes and the attenuation along the street is modelized with $m+2$ image sources (see eq. (\ref{eq:source_image})).

\begin{figure}[h!]
\begin{center}
\includegraphics[width=12cm]{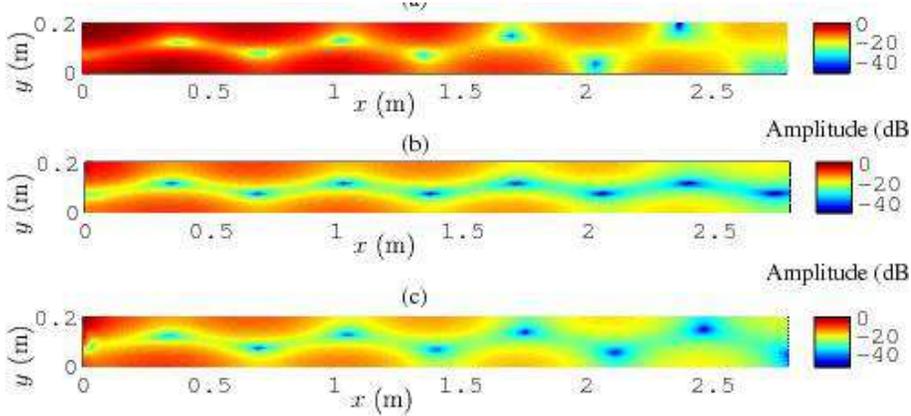}
\end{center}
\caption{(a) Experimental acoustic pressure in a street (scale model) for $f=1000$ Hz at $z=0.07$ m and for a square source centered on $y_s=0.175$ m and $z_s=0.07$ m, (b) analytical acoustic pressure in a street at $z=0.07$ m, for $f=1000$ Hz and for a square source centered on $y_s=0.175$ m., (c) adaptated 2D-FDTD simulation of the acoustic pressure in the $2$D-street for $f=1000$ Hz and for a square source centered on $y_s=0.175$ m.}
\label{fig:comp1000}
\end{figure}

The fig. \ref{fig:comp1000}b proposes the analytical pressure map in the street computed by means of the eq. (\ref{eq:pressure_total}) with $N=56$, $m=3$  and $R_r=-0.3$ for $f=1000$ Hz with a square source centered on $y_s=0.175$ m and at the altitude $z=0.07$ m. The qualitative agreement with the corresponding experimental result (see fig. \ref{fig:comp1000}a) is correct : the analytical shape of the field is close to the experimental one. For this case, at the end of the street, the location on the $y$-axis of the minimum of the analytical map are shifted refering to the experimental results. This difference can be explained by a default in the alignement of the robot and the street $x$-axis. 

Fig. \ref{fig:comp1000}c shows the simulation of the acoustic pressure along the street for $f=1000$ Hz with a square source centered on $y_s=0.175$ m. The step of spatial sampling of the FDTD is $0.01$ m. The comparison between fig. \ref{fig:comp1000}c and \ref{fig:comp1000}a shows a good agreement between both results. The simulated result exhibits the same shape of the acoustic field than the experimental one. The attenuation of the simulated and experimental results are in the same order which prooves that, for this frequency, the acoustic radiation conditions for the street roof can be modelized by an attenuation coefficient. 

\begin{figure}[h!]
\begin{center}
\includegraphics[width=12cm]{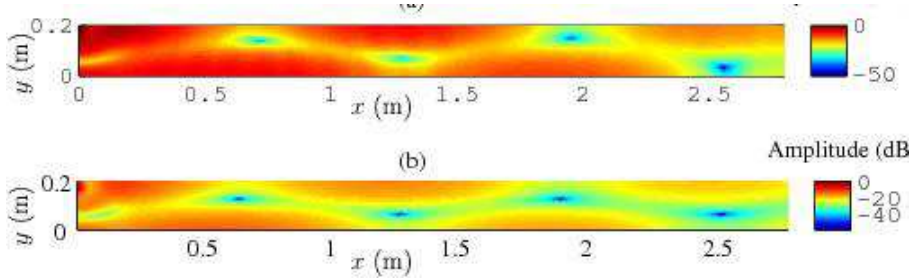}
\end{center}
\caption{(a) Acoustic pressure in the street (scale model) for $f=1500$ Hz at $z=0.07$ m and for a square source centered on $y_s=0.175$ m and $z_s=0.07$ m, (b) analytical pressure map in a street at $z=0.07$ m for $f=1500$ Hz and for a square source centered on $y_s=0.175$ m.}
\label{fig:comp1500}
\end{figure}

The fig. \ref{fig:comp1500}b shows the analytical calculation of the acoustic pressure in the street at $z=0.07$ m for a frequency $f=1500$ Hz with the same source condition than in the experimental study. The number of modes is $N=59$, the value of the reflexion coefficient is estimated to $R_r=-0.1$ and we use $5$ image sources. The qualitative comparison of this result with the experimental one (see fig. \ref{fig:comp1500}a) shows a good agreement except at the end where the analytical model overestimates pressure compared to experiments.

\begin{figure}[h!]
\begin{center}
\includegraphics[width=12cm]{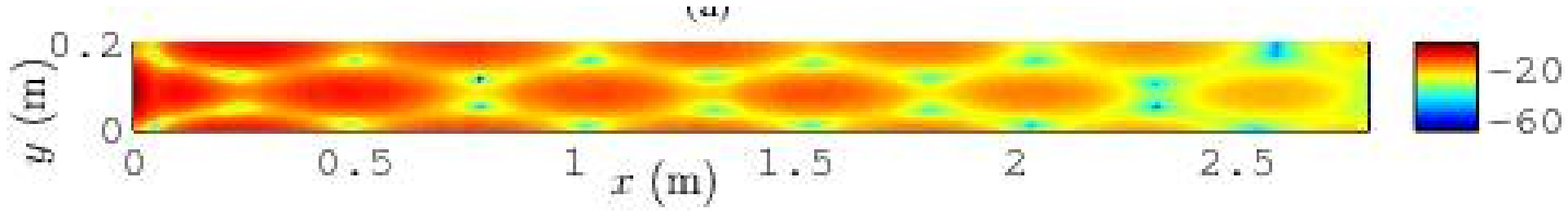}
\includegraphics[width=12cm]{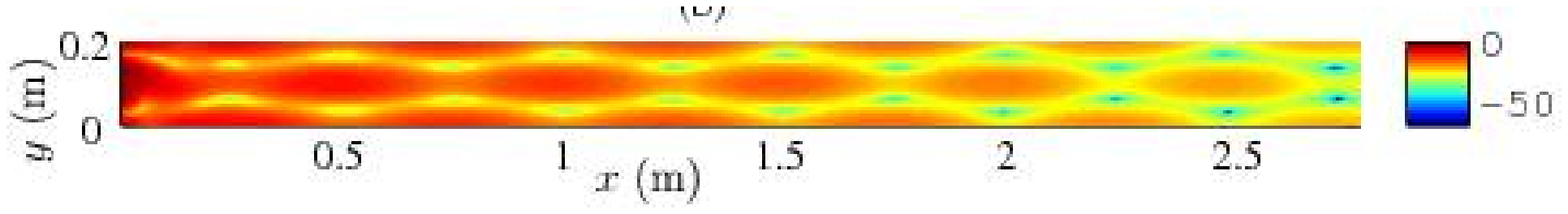}
\end{center}
\caption{(a) Acoustic pressure in the street (scale model) for $f=2500$ Hz at $z=0.07$ cm and for a square source centered on $y_s=0.1$ m and $z_s=0.07$, (b) adaptated $2$D-FDTD simulation of the acoustic pressure in the $2$D-street for $f=2500$ Hz and for a point source centered on $y_s=0.1$ m.}
\label{fig:comp2500}
\end{figure}

The fig. \ref{fig:comp2500} proposes the simulated map of the acoustic pressure along the street for $f=2500$ Hz. The step of spatial sampling of the FDTD is $0.01$ m. The comparison of this result with the experimental map shows a good agreement. The mode distribution of the pressure field along the street is the same for the experimental and simulated results. The acoustic radiation conditions are taken into account with qualitatively good precision. 

In view of these primary results, the effect of the street open roof can be modelized by a attenuation coefficient in the $2$D adapted FDTD computation depending on the frequency : for $f=1000$ Hz, $\alpha=0.0018$ s$^{-1}$ and for $f=2500$ Hz, $\alpha=0.0025$ s$^{-1}$. The attenuation is greater at $2500$ Hz than at $1000$ Hz. For the analytical model, the open roof effect is described by a negative reflexion coefficient with a amplitude which decreases when the frequency increases : $R_r=-0.3$ for $f=1000$ Hz, $R_r=-0.1$ for $f=1500$ Hz and $R_r=0$ for $f=2000$ Hz and $2500$ Hz. This result shows that more the frequency increases, more the open roof radiation conditions are close to a complete absorption.

\subsection{Quantitative comparison : low frequency validation}

\begin{figure}[h!]
\begin{center}
\includegraphics[width=13cm]{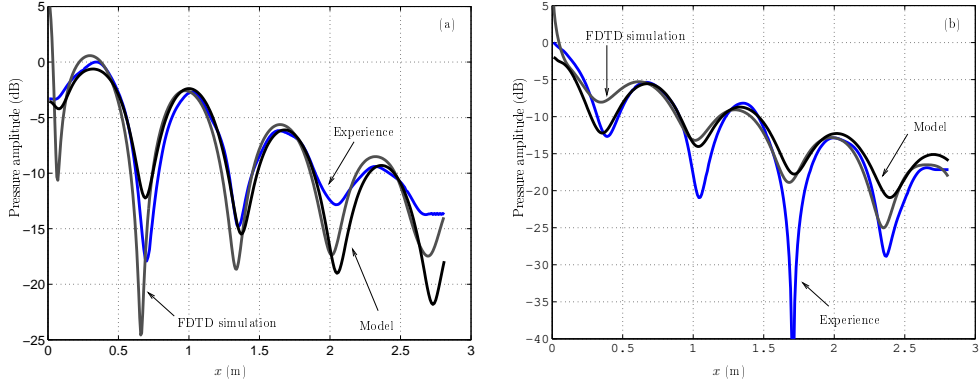}
\end{center}
\caption{Comparison of the experimental (blue curve), analytical (black curve) and numerical (gray curve) results of the pressure profile along the street for $f=1000$ Hz at $y=0.09$ m (a) and for $f=1000$ Hz at $y=0.15$ m (b).}
\label{fig:comp_Exp_Model_Simul_1000}
\end{figure}

To study with more precision the differences between the experimental, modelized and simulated results, the acoustic pressure along the street for $y=0.09$ m and $y=0.15$ m is compared for $f=1000$ Hz on figs. \ref{fig:comp_Exp_Model_Simul_1000}a and \ref{fig:comp_Exp_Model_Simul_1000}b. The analytical model proposes a good prediction of the pressure attenuation and of the maximum pressure positions along the street. With the adaptated $2$D-FDTD simulation, the attenuation is well predicted and the maxima positions of the pressure field along the street are estimated with a maximum error of $0.05$ m. This result prooves that a reflexion coefficient used in the analytical model and a negative pressure source uniformly distributed along the street used in the 2D-FDTD simulation constitute a good modelization for the pressure attenuation along the street due to the open roof. 

\begin{figure}[h!]
\begin{center}
\includegraphics[width=8cm]{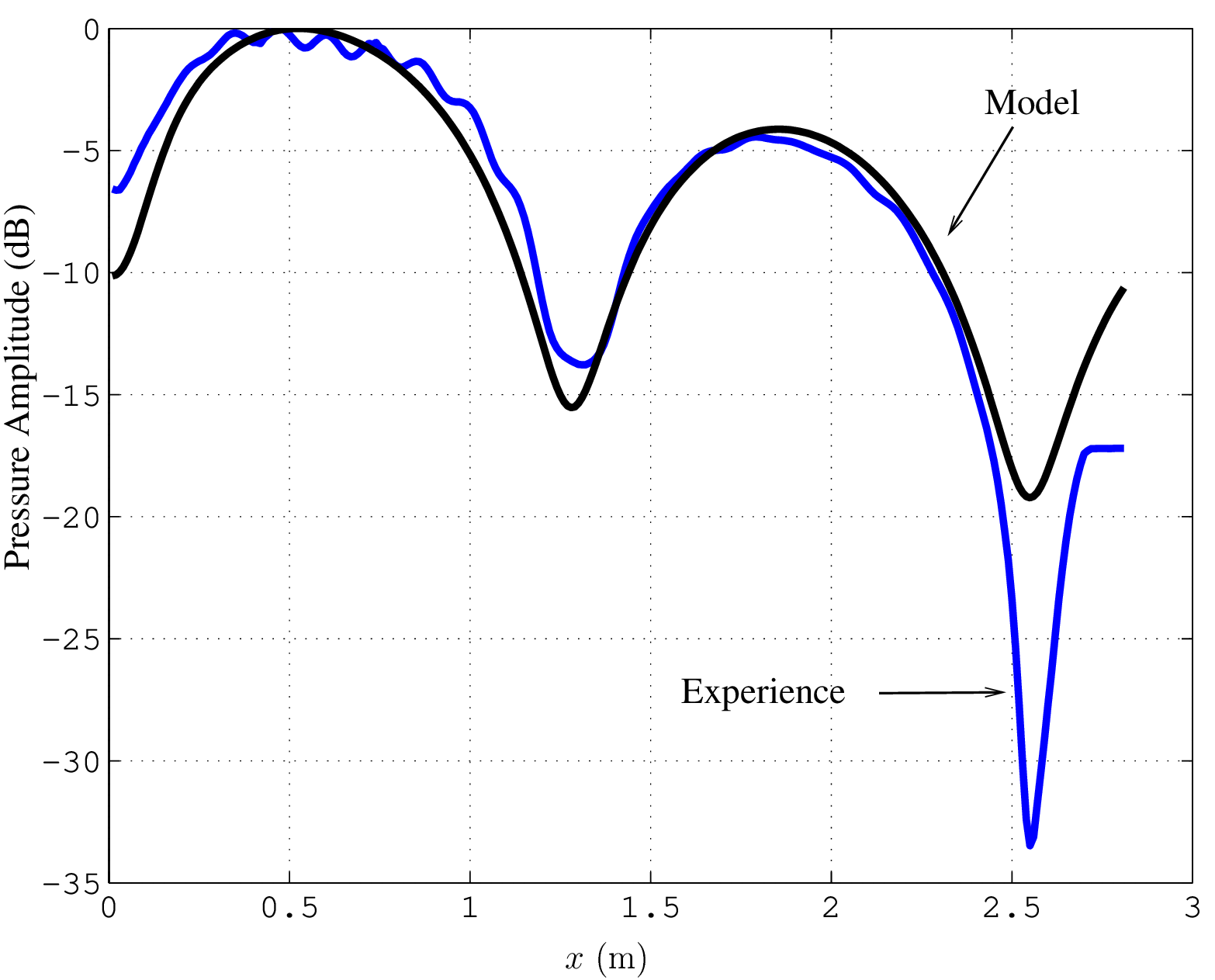}
\end{center}
\caption{Comparison of the experimental (blue curve) and analytical (black curve) results of the pressure profile along the street for $f=1500$ Hz at $y=0.05$ m.}
\label{fig:comp_Exp_Model_1500}
\end{figure}

The experimental and analytical acoustic pressure profiles along the street for $f=1500$ Hz at $y=0.05$ m are proposed in fig. \ref{fig:comp_Exp_Model_1500}. This comparison shows a good agreement between the analytical model and the experimental study. The maximum locations of the pressure along the street are well estimated and the global attenuation is predicted with a good accuracy. This prooves that the use of a negative reflexion coefficient can be a good modelization of the acoustic radiation conditions by the open roof at this frequency. 

%-------------------------------------------------------
\subsection{Frequency limit of the modelization}
\label{limit}

\begin{figure}[h!]
\begin{center}
\includegraphics[width=8cm]{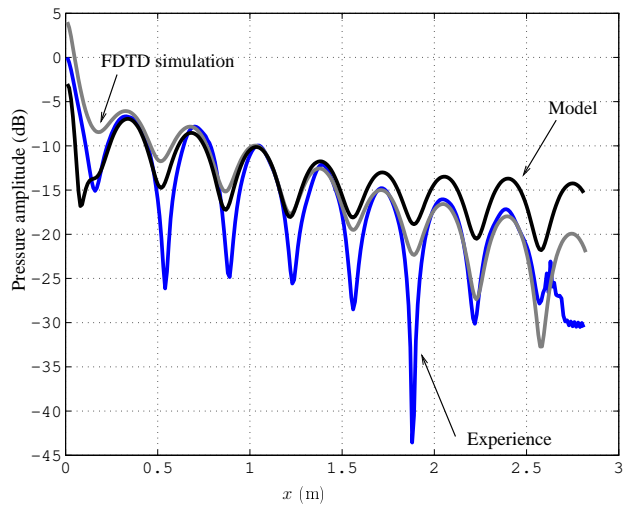}
\end{center}
\caption{Comparison of the experimental (blue curve), analytical (black curve) and numerical (gray curve) results for the pressure attenuation along the street for $f=2000$ Hz at $y=0.08$ m.}
\label{fig:comp_Exp_Model_Simul_2000}
\end{figure}

For $f=2000$ Hz, the analytical, simulated and experimental results are compared on the fig. \ref{fig:comp_Exp_Model_Simul_2000} for $y=0.08$ m. The analytical model is used with $R_r=0$ and $N=62$ and the simulated one is performed with $\alpha=0.0020$. The agreement between the simulated and experimental curves is good : the attenuation is very well predicted and a error of $0.03$ m on the maximum positions along the street, involving notably by the discretization of the FDTD computation, is visible. 

For the analytical results at $f=2000$ Hz, the limits of the model are reached. The positions of the maximum are well estimated with this method but the attenuation is badly predicted. The reflexion coefficient is chosen to zero which involves a maximum for the pressure attenuation. But the comparison with the experimental data shows that the analytical attenuation is underestimated : at $x=2.5$ m the error between the model and the experience for the acoustic pressure is greater than $3$ dB.

\begin{figure}[h!]
\begin{center}
\includegraphics[width=8cm]{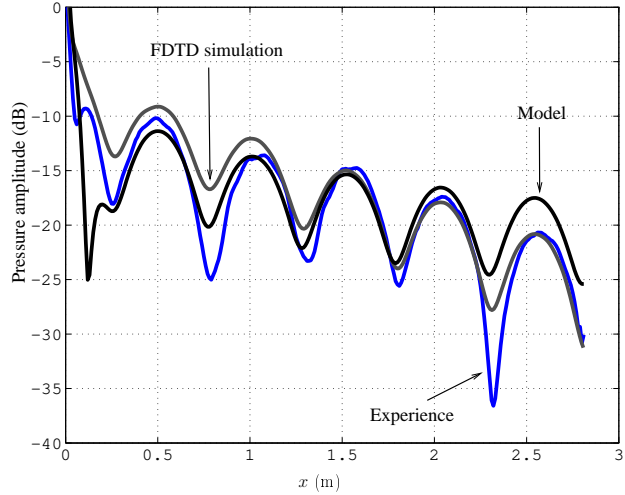}
\end{center}
\caption{Comparison of the experimental (blue curve), analytical (black curve) and numerical (gray curve) results of the pressure profile along the street for $f=2500$ Hz at $y=0.1$ m.}
\label{fig:comp_Exp_Model_Simul_2500}
\end{figure}

The fig. \ref{fig:comp_Exp_Model_Simul_2500} shows the experimental, analytical and simulated acoustic pressure along the street for $f=2500$ Hz at $y=0.1$ m. The simulated result is obtained with $\alpha=0.025$ and the analytical model is performed for $R_r=0$ with $11$ image sources (corresponding to $m=5$). As for $f=2000$ Hz, the agreement between the numerical and experimental results is good : the pressure attenuation and the maximum positions of the pressure field are well predicted excepted the near foeld of the source. The analytical model overestimates the pressure attenuation along the street but proposes a good estimation of the maximum positions.

\begin{figure}[h!]
\begin{center}
\includegraphics[width=12cm]{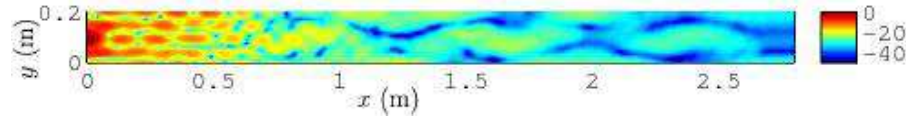}
\end{center}
\caption{Experimental acoustic pressure in a street at $z=0.07$ m, for a frequency $f=3400$ Hz and for a square source centered on $y_s=0.1$ m and $z_s=0.07$ m.}
\label{fig:mesure3400}
\end{figure}

The analytical model or the numerical simulation, developped in this work, are based on a low frequency hypothesys which permits to modelize the attenuation by a single coefficient depending on the frequency and the street size. By increasing the frequency, the acoustic radiation losses, due to the open roof of the street, is not anymore uniformly distributed on all modes in the waveguide. The notion of leaky modes \cite{Pelat2008} (with its own attenuation or leakage) must be introduced in the modelization to describe the acoustic propagation along the street. The fig. \ref{fig:mesure3400} shows the map of the experimental pressure amplitude along the street for a frequency $f=3400$ Hz. 

\begin{figure}[h!]
\begin{center}
\includegraphics[width=8cm]{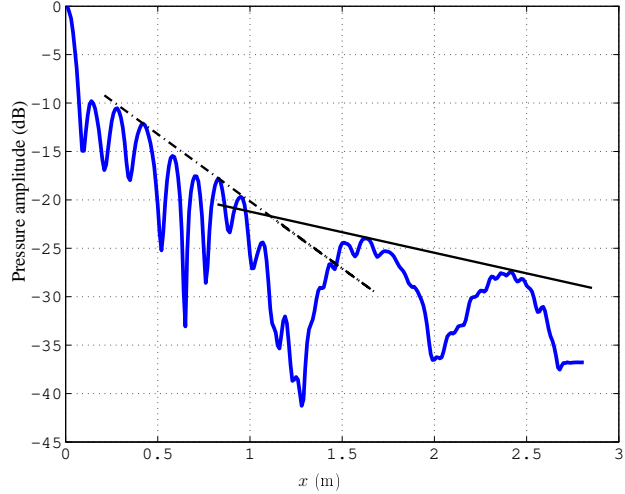}
\end{center}
\caption{Experimental pressure profile along the street for $f=3400$ Hz at $y=0.1$ m.}
\label{fig:mesure3400_y10cm}
\end{figure}

On the fig. \ref{fig:mesure3400_y10cm}, the pressure profile for $f=3400$ Hz at $y=0.1$ m is represented. The difference of leakage between the modes is easily shown at the beginning of the street : after $1$ m along the street, a leaky mode (defined by $2 \lambda = d$ along the $y$-axis) disappears completely with a global attenuation illustrated by the shaded line while a second leaky mode (defined by $\lambda/2=d$ along the $y$-axis) propagates with a different attenuation illustrated by the second line.

%%%%%%%%%%%%%%%%%%%%%%%%%%%%%%%%%%%%%%%%%%%%%%%%%%%%%%%%%%%%%%%%%%%%%%%%%%%%%%%%%%%%%%%%%%%%%%%%%
\section{Conclusion}
\label{conclusion}

In this work, we have shown that it is possible to modelize and to simulate the acoustic propagation along an urban street with a good agreement comparing to experimental results. For the low frequency case (until $100$ Hz at full scale), an adaptated $2$D-FDTD simulation, taking into account the acoustic losses due to the radiations by the street roof, proposes a good modelization of the acoustic propagation. A coefficient $\alpha$ of the attenuation along the street is adapted by fitting with the experimental results.\\
An analytical model for the acoustic propagation along a street canyon is also developped by mixing a modal approach to describe the propagation in the horizontal plane and a $2$D propagation model using image sources to modelize the attenuation along the street. A reflexion coefficient is defined to modelize the role of the street open roof. For the low frequency case, the decrease of the reflexion coefficient modulus with the increase of the frequency shows that the acoustic radiation conditions of the open roof depend on frequency and prooves that more frequency increases less reflexions on the open roof are present. For higher frequencies, while the adaptated $2$D-FDTD simulation describes well the pressure field, the analytical model is limited by the presence of numerous leaky modes with different attenuations.

\section*{Acknowledgements}

The authors thank S. Lebon, P. Collas and P.E. Chartrain for their help on the experimental system, post-processing computation and experimental study.

\end{document}